\documentclass[9pt, conference]{IEEEtran}
\usepackage{hyperref}
\usepackage{setspace}
\IEEEoverridecommandlockouts
\usepackage{cite}
\usepackage{amsmath,amssymb,amsfonts}
\usepackage{graphicx}
\usepackage{textcomp}
\usepackage{xcolor}
\usepackage[noend]{algpseudocode}
\usepackage{algorithm}
\usepackage{amsmath}
\usepackage{amssymb}
\usepackage{amsfonts}
\usepackage{tikz}
\usepackage{booktabs}
\usepackage{multirow}
\usepackage{graphicx}         
\usepackage{color}
\usepackage{cite}             
\usepackage{comment}          
\usepackage{soul}             
\soulregister\cite7
\soulregister\ref7
\soulregister\pageref7
\usepackage{amsthm}
\usepackage{etoolbox}         
\usepackage{url}
\usepackage{nth}              
\usepackage{bm}               
\usepackage{caption} 
\usepackage{subfig}

\usepackage{cleveref}                                      

\theoremstyle{plain}

\newtheorem*{theorem*}{\textbf{Theorem}}

\theoremstyle{definition}

\newtheorem*{problem*}{\textbf{Problem}}

\makeatletter
\let\OldStatex\Statex
\renewcommand{\Statex}[1][3]{%
  \setlength\@tempdima{\algorithmicindent}%
  \OldStatex\hskip\dimexpr#1\@tempdima\relax
}
\makeatother

\graphicspath{{./figs/}}
\makeatletter
\newcommand{\linebreakand}{%
  \end{@IEEEauthorhalign}
  \hfill\mbox{}\par
  \mbox{}\hfill\begin{@IEEEauthorhalign}
}
\makeatother

\usepackage{color}
\usepackage{pifont}
\usepackage{CJKutf8}

\begin{document}



\title{
  \huge{
  \textbf{
AnalogXpert: Automating Analog Topology Synthesis by Incorporating Circuit Design Expertise into Large Language Models
  }%
  }
  \thanks{This work is supported in part (authors from Peking University) by National Science and Technology Major Project (2021ZD0114702), the Natural Science Foundation of Beijing, China (Grant No.Z230002), project QYJS-2023-2303-B, and the 111 project (B18001).}
  \vspace{-0.5cm}
}%

\author{
\large
Haoyi Zhang$^1$, 
Shizhao Sun$^{4*}$,
Yibo Lin$^{1,2,3*}$, 
Runsheng Wang$^{1,2,3}$, 
Jiang Bian$^4$ \\
\small
$^1$School of Integrated Circuits, Peking University, Beijing, China 
$^2$Institute of EDA, Peking University, Wuxi, China \\
$^3$Beijing Advanced Innovation Center for Integrated Circuits, Beijing, China
$^4$Microsoft Research Asia, Beijing, China \\
$^{*}$Corresponding author: Yibo Lin (yibolin@pku.edu.cn), Shizhao Sun (shizsu@microsoft.com)
\vspace{-0.4cm}
}

\maketitle

\begin{abstract} 
Analog topology synthesis is one of the major challenges in analog design automation since the topology of analog circuits has a large design space and contains a lot of human expertise. Traditional methods suffer in generating high-quality topology due to the diversity of topologies and the lack of ability to understand human experience.  Therefore, LLM has been adopted in recent studies to generate such topologies. However, most of the existing work utilizes ideal model-based generation or ambiguous design requirements, both of which are not in line with industrial practice and require additional effort. In this work, we propose AnalogXpert, an LLM-based agent formulating topology synthesis as subcircuit-level SPICE code generation which is more practical. AnalogXpert incorporates circuit design expertise by introducing a proofreading strategy that allows LLMs to incrementally correct the errors in the initial design. Finally, we construct a high-quality benchmark validated by both real data (30) and synthetic data (2k). AnalogXpert achieves 40\% and 23\% success rates on the synthetic dataset and real dataset respectively, which is markedly better than those of GPT-4o (3\%,3\%) and AnalogCoder (8\%,6\%).

\end{abstract}

\section{Introduction}

Analog circuits~\cite{analog1,analog2} form the backbone of many modern electronic systems, playing a crucial role in processing continuous signals to achieve a variety of tasks in the devices that permeate our everyday lives. Analog circuit design usually can be divided into three stages: \textbf{(1) Topology synthesis}~\cite{zhaoDeepReinforcementLearning2022}. Topology synthesis determines the whole analog circuit topology. This stage will select the basic devices (e.g. transistors, capacitors, resistors) and give the connection relationship among these basic devices. \textbf{(2) Circuit Sizing}~\cite{sizing1,sizing2}.  The selected basic devices need to be applied with the appropriate parameters to maximize the circuit performance for a given topology. \textbf{(3) Layout synthesis}~\cite{chenMAGICALOpenSource2021,zhangSAGERouteHierarchicalAnalog,zhangSAGERouteSynergisticAnalog,dharALIGNOpensourceAnalog2020a}. This stage performs the placement and routing for an analog circuit to generate the final layout.

Topology synthesis is the foundation of the entire circuit design and determines the performance of the circuit. Topology design of analog circuits heavily relies on designers’ expertise and experience, which largely determines the circuit performance after tape-out. As design experts require long training periods and are very expensive, automating topology design becomes urgently desired to push cutting-edge analog circuits. Some research has emerged and focused on the automation of it as depicted in Table~\ref{method_comparison}. Before the introduction of LLMs, some research has already attempted to automate analog topology synthesis with versatile AI methods. For example, RLATS~\cite{zhaoDeepReinforcementLearning2022} leverages reinforcement learning (RL) to build up the analog circuit step by step. CKTGNN~\cite{dongCktGNNCircuitGraph2024a} builds up a synthetic dataset of ideal models to train a graph neural network that achieves a decent performance. Ideal models usually replace the combination of several transistors with a single transconductance. For example, a five-transistor OTA can be represented by a single transconductance. These methods lack the ability to understand human experience and struggle to effectively generate reasonable analog circuit topologies. 

With the emergence and continuous improvement of large language model (LLM) technology~\cite{wei2023chainofthoughtpromptingelicitsreasoning,yao2023treethoughtsdeliberateproblem}, some research leverages the LLM to generate the analog topology. LADAC~\cite{liuLADACLargeLanguage2024} and AnalogCoder\cite{laiAnalogCoderAnalogCircuit2024} leverage prompting to enhance the analog design ability of LLM, respectively. LaMAGIC~\cite{changLaMAGICLanguageModelbasedTopology} builds up a synthetic dataset to support the SFT of LLM, but its design tasks are focused on the power converters which are very different from conventional analog circuits such as OTA, LDO, and so on. Although these studies make valuable exploration, there is still a gap between the scenario they focused on and the practical application scenarios due to the following reasons: (1) The design requirements are very ambiguous; (2) The gap between ideal models and a real analog circuit is huge; (3) The benchmarks in previous work contain either very few real cases or only synthetic cases.

\begin{figure}[tb]
    \centering
    \includegraphics[width=0.48\textwidth]{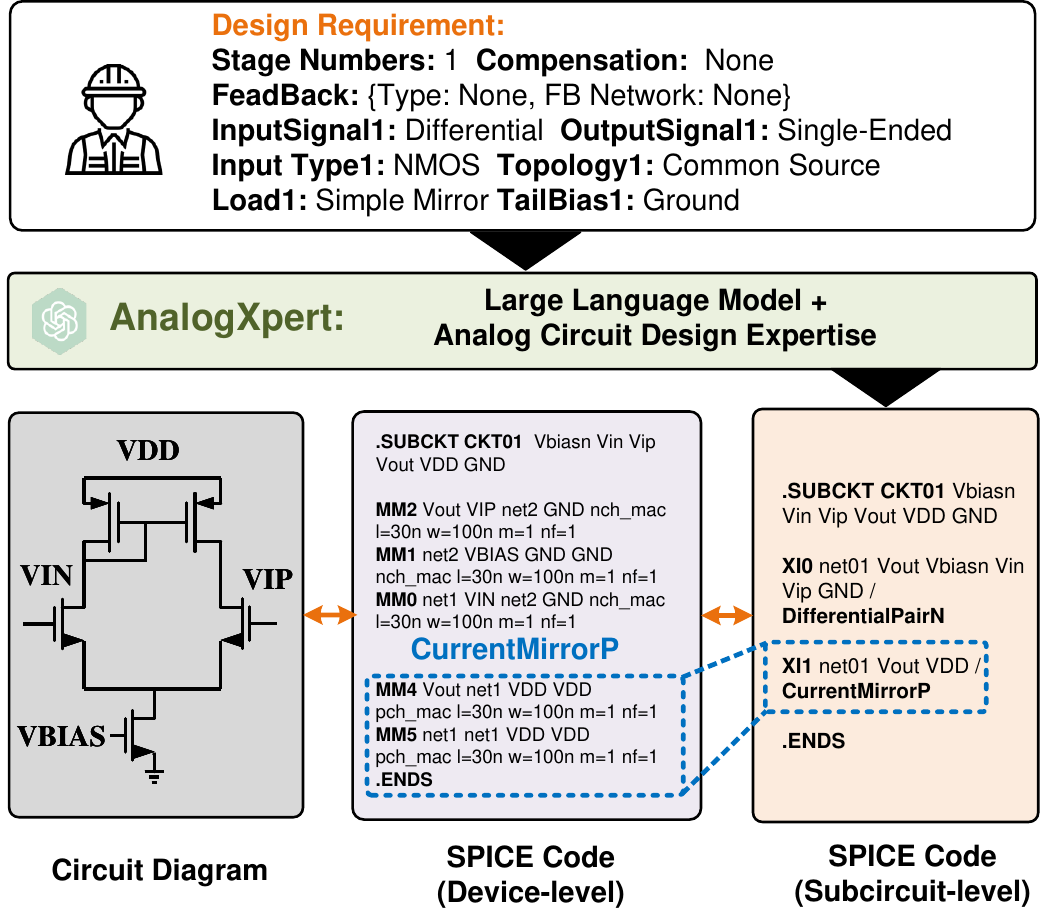}
    \caption{AnalogXpert formulation of the topology synthesis task.}
    \label{intro}
    \vspace{-0.5cm}
\end{figure}

In this work, we introduce AnalogXpert, designed to address more practical analog topology synthesis problems while utilizing circuit design expertise for better performance. As shown in Figure~\ref{intro}, AnalogXpert takes detailed design requirements of structure as input and outputs a real circuit topology design in subcircuit-level SPICE code format (rather than ideal models), which is a conditional generation task. Based on the above formulation, AnalogXpert presents an LLM-based agent for topology synthesis by incorporating circuit design expertise into LLMs. First, a well-designed subcircuit library is proposed to reduce the design space and improve the generation success rate.  Second, the design task decomposition is performed to make the design more logical and easy to check for errors. AnalogXpert leverages the Chain-of-Thought (CoT) to prompt LLMs to generate topology step by step. Third, AnalogXpert proposes a proofreading strategy to check the generated analog topologies and give revision messages back to LLMs for iterative refinement.

\begin{table}[bt]
        \caption{\small{Comparison of AnalogXpert and related works.
}}
\label{method_comparison}
\resizebox{0.48\textwidth}{!}{
\begin{tabular}{l|cccc}
\toprule
\multirow{2}{*}{\textbf{Method}}  & \textbf{Design Space} & \textbf{Benchmark} & \textbf{Use} & \textbf{Refinement} \\  
  & \textbf{Reduction Method}\textsuperscript{1} & \textbf{Type} & \textbf{LLM} & \textbf{Method}\\  
\midrule
CKTGNN~\cite{dongCktGNNCircuitGraph2024a}   & Ideal Model  & synthetic & x & -- \\ 
RLATS~\cite{zhaoDeepReinforcementLearning2022}   & Subcircuit Library & real & x & --\\ 
\midrule
LADAC~\cite{liuLADACLargeLanguage2024} & -- & real & \checkmark & --  \\
LaMAGIC~\cite{changLaMAGICLanguageModelbasedTopology}  & Ideal Model & synthetic & \checkmark  & -- \\
Artisan~\cite{chen24artisan} & Ideal Model & real &  \checkmark & --  \\
AnalogCoder~\cite{laiAnalogCoderAnalogCircuit2024} & -- & real & \checkmark & simulation feedback \\ 
\midrule
\multirow{2}{*}{\textbf{AnalogXpert}}  & \multirow{2}{*}{Subcircuit Library} & \textbf{real 
 \&} & \multirow{2}{*}{\checkmark} & \textbf{human expertise} \\ 
 &  & \textbf{synthetic} &  & \textbf{based proofreading} \\ 

\bottomrule
\end{tabular}
}
{\scriptsize
\textsuperscript{1} Method to simplify the analog circuits design task. 
}
\vspace{-0.5cm}
\end{table}



Since the task formulation of AnalogXpert is very different from previous works, we construct a new benchmark including both real cases (30) and synthetic cases (2k). Each case includes some specific design requirements that the generation task should honor. Previous studies only have one type of data, either real data or synthetic data. Meanwhile, the number of real data is very limited ($\leq$5), except for the AnalogCoder~\cite{laiAnalogCoderAnalogCircuit2024} (24). Validated on the proposed benchmark, AnalogXpert achieves 40\% and 23\%
success rates in the synthetic dataset and real dataset respectively, which is much
better than the GPT-4o (3\% in the synthetic dataset, 3\% in the real dataset) and AnalogCoder~\cite{laiAnalogCoderAnalogCircuit2024} (8\% in the synthetic dataset, 6\% in the real dataset). The experimental results demonstrate the effectiveness and robustness of the AnalogXpert.

The main \textbf{contributions} of this paper can be summarized as follows: 

\begin{itemize}
    \item  We focus on a more practical topological synthesis problem with detailed inputs which is a conditional generation task. We formulate it as a subcircuit-level SPICE code generation problem by introducing a design space reduction method based on an extensible subcircuit library.
    \item We propose a CoT-based LLM agent to imitate the human design process, decomposing topology synthesis tasks into subcircuit block selection and connection graph construction. 
    \item We propose a proofreading strategy based on the human experience, which makes LLMs revise the generated topology iteratively, to further improve the design ability of LLM agents. With several rounds of self-refinement, LLM agents can avoid some basic mistakes and improve the design success rate. 
    \item We also propose a holistic benchmark to completely validate the design ability of AnalogXpert. The proposed benchmark consists of 30 real-world analog design tasks and 2k synthetic data. 
\end{itemize}



\section{Preliminaries}

\subsection{Analog Topology Synthesis Automation}
Analog topology synthesis is the most challenging step in the analog design flow and thus has attracted extensive research interest. Before the introduction of LLMs, some research has already attempted to automate analog topology synthesis with versatile AI methods. For example, RLATS~\cite{zhaoDeepReinforcementLearning2022} leverages reinforcement learning (RL) to build up the analog circuit step by step. RLATS establishes a subcircuit library to simplify the topology synthesis problem so that the RL agent is able to handle it. However, the design diversity of analog circuits makes it difficult to transfer RL agents between different circuit types. CKTGNN~\cite{dongCktGNNCircuitGraph2024a} builds up a 10k dataset to train a graph neural network which achieves an impressive performance. The drawback is that CKTGNN leverages ideal models to reduce the design space. Ideal models can not represent the entire legalized design space and can not be directly converted to the final topology. With the introduction of LLM, more research works on the automation of topology synthesis have emerged. LADAC~\cite{liuLADACLargeLanguage2024} and Artisan~\cite{chen24artisan} leverage prompting and supervised fine-tuning (SFT) to enhance the analog design ability of LLM, respectively. However, they only validate the proposed framework on a few ($\leq$5) real analog cases. This is not enough to demonstrate that LLM has adequate analog circuit design capabilities. LaMAGIC~\cite{changLaMAGICLanguageModelbasedTopology} build up a 120k synthetic dataset to support the SFT of LLM, but its design tasks are simpler than the actual analog design tasks. LaMAGIC actually focuses on a kind of radio frequency circuit with limited basic devices ($\leq$ 6) including a capacitor (2 terminals), inductor (2 terminals), and switch (2 terminals). Analog usually is made up of tens of basic devices (0$\sim $50), such as transistors (4 terminals), capacitors (2 terminals), and resistors (2 terminals). AnalogCoder\cite{laiAnalogCoderAnalogCircuit2024} leverages a prompt-based LLM to generate the analog topology and validate the framework on some real data. The disadvantage is that user requirements are very ambiguous and can be handled directly by existing LLM agents in most cases. In real-world scenarios, designers would like to give more detailed design conditions. Therefore, we propose AnalogXpert to process concrete conditions and validate the framework on both real data and synthetic data which is a more comprehensive demonstration of LLM's analog design capabilities.   

\subsection{Large Language Models} 
Large language models with pre-trained parameters such as GPT, LLaMa~\cite{dubey2024llama3herdmodels}, Claude, have demonstrated terrific ability in versatile tasks. Recent research on prompting has further enhanced the LLM ability. For example, CoT~\cite{wei2023chainofthoughtpromptingelicitsreasoning} is a technique that encourages language models to generate intermediate reasoning steps in a step-by-step manner, rather than directly providing the final answer. This method improves the model's ability to solve complex tasks by making its reasoning process more transparent and structured. Some other researches focus on In-context learning~\cite{dong2024surveyincontextlearning} which involves giving a language model examples of tasks or instructions within the same prompt to improve the final performance. In addition to pre-generation promptings, post-generation feedback is equally important. Some feedback techniques such as self-reflection~\cite{renze2024selfreflectionllmagentseffects}, can also improve the LLM performance. These techniques provide the foundation for constructing agents with practical functions, such as gene-editing~\cite{huangCRISPRGPTLLMAgent2024a}, math~\cite{ahn2024largelanguagemodelsmathematical}, and chip designs~\cite{liuChipNeMoDomainAdaptedLLMs2024}. Based on these techniques, we incorporate the circuit design expertise into LLMs, which allows LLM to really deal with analog topology synthesis problem.

\section{AnalogXpert Framework}

\begin{figure*}[tb]
    \centering
    \includegraphics[width=0.9\textwidth]{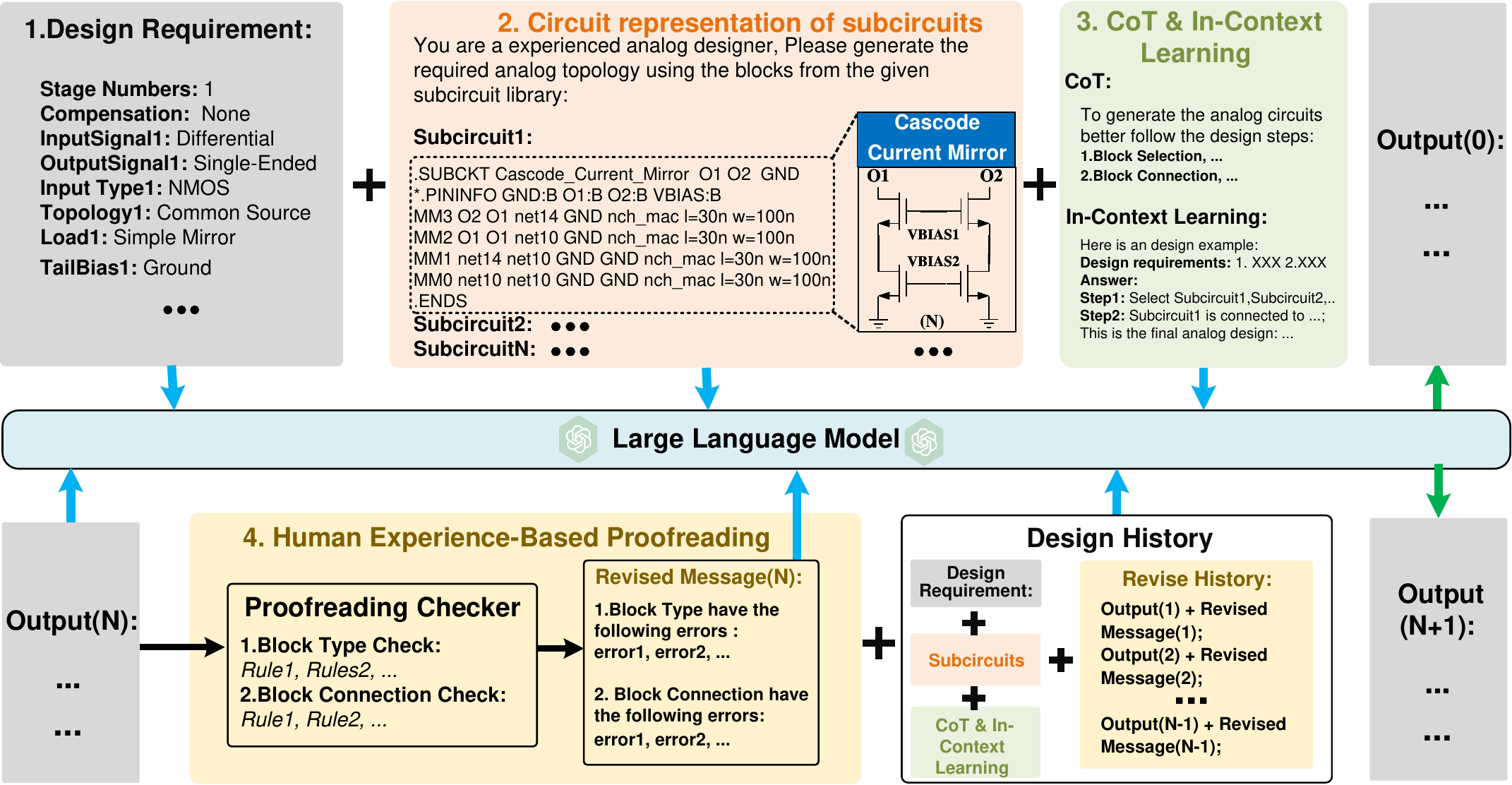}
    \caption{Overview of AnalogXpert framework.}
    \label{Overview}
    \vspace{-0.5cm}
\end{figure*}
In this section, we elaborate on the detailed methods of AnalogXpert (as Figure~\ref{Overview} shows), an efficient training-free analog design agent incorporating human design expertise. AnalogXpert introduces a brand new task formulation of conditional analog topology synthesis which can demonstrate the design ability of LLM agents (Section~\ref{Problem}). Leveraging a well-designed subcircuit library, AnalogXpert constructs a novel and effective analog circuit representation which greatly helps LLM agents design analog circuits concisely (Section~\ref{Subcircuit}). A domain-specific prompting design flow based on CoT is proposed and further enhanced by the in-context learning (Section~\ref{prompting}). Although the methods mentioned above can improve the design capability of LLM agents, the single-round generation mode still struggles to handle complex tasks. Therefore, AnalogXpert further introduces human experience-based proofreading to help LLM agents gradually correct the mistakes and finally generate the correct topology in several rounds (Section~\ref{selfrefinement}). With the cooperation of these techniques, AnalogXpert has a decent performance in the conditional analog topology synthesis tasks.

The overview of the AnalogXpert framework is shown in Figure~\ref{Overview}. AnalogXpert takes versatile design requirements of analog circuit structures as input, such as stage number, input signal type, feedback type, and so on. AnalogXpert performs the topology synthesis based on given subcircuit blocks that are in SPICE code format for easy use by LLM agents. The proposed CoT first selects the subcircuit blocks and then determines block connection relationships. Meanwhile, a corresponding design example is provided for the in-context learning. After obtaining the initial results, the circuit design will be verified by a proofreading checker in terms of block types and block connections. If there is an error in the circuit design then a revised message will be generated by the checker. The revised message as well as the previous generation history are given to the LLM agents and the next result will be generated.

\subsection{Problem Formulation}
\label{Problem}
Previous analog topology synthesis usually formulates the analog topology synthesis problem as Equation~\ref{traditiondef} shows. Users give the required circuit type $Ckt$, and a series of required design specifications $\sum_{n=1}^NSpec_n$. For example, the bandwidth and phase margin are two typical specifications of an amplifier. The automatic tool is performed as the function $F\{\}$ and generates the final analog circuits. A typical analog circuit topology consists of the selected devices $\{D_i\}$ and the connection relationship between them $\{\sum_{n=1}^NN_j(D_i,T_k)_n\}$, where $T_k$ denotes the $k$ terminal of the devices. However, there are two disadvantages of this formulation: \textbf{(1)The input condition~($\sum_{n=1}^NSpec_n$) is ambiguous.} A series of required design specifications can be satisfied with many different topologies.  At the same time, one topology may also satisfy different specifications. Therefore, using design specifications as input conditions is too ambiguous for a generation task. \textbf{(2)The input condition~($\sum_{n=1}^NSpec_n$) includes complex mathematical calculations.}  The relationship between the analog topology and design specifications is often described by some complex mathematical equations which is difficult and inappropriate for LLM agents to deal with. 

\begin{equation}
    \label{traditiondef}
    \{D_i\},\{\sum_{n=1}^NN_j(D_i,T_k)_n\} = F\{Ckt,\sum_{n=1}^NSpec_n\}
\end{equation}

To this end, we propose a brand new formulation of analog topology synthesis as Equation~\ref{newdef} shows. The difference lies in the input conditions. AnalogXpert leverages some structure requirements $\sum_{n=1}^NStruc_n$ to instead the specifications $\sum_{n=1}^NSpec_n$. Structure requirements directly describe the characteristics of the analog topology which are more concrete than the specifications. Meanwhile, structure requirements have successfully separated the complex mathematical calculations from the generation tasks. Such formulation turns the analog topology synthesis problem into a pure sequence-to-sequence problem which is more appropriate for LLM agents.    

\begin{equation}
    \label{newdef}
    \{D_i\},\{\sum_{n=1}^NN_j(D_i,T_k)_n\} = F\{Ckt,\sum_{n=1}^NStruc_n\}
\end{equation}


\begin{figure*}[tb]
    \centering
    \includegraphics[width=0.9\textwidth]{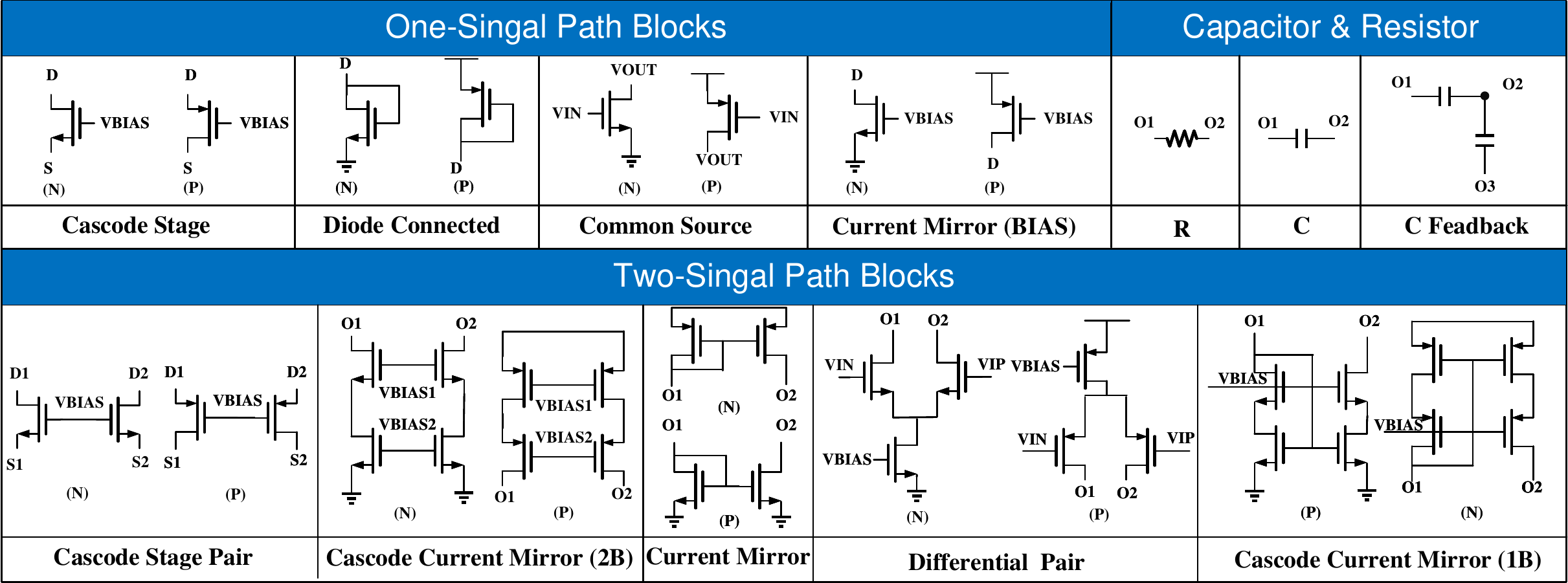}
    \caption{Details of subcircuit library and subcircuit-level SPICE code representation for analog topology.}
    \label{subcircuit}
    \vspace{-0.5cm}
\end{figure*}


\subsection{Analog Circuit Representation}
\label{Subcircuit}
The conventional SPICE code is built up from devices, such an approach leads to a flexible but complex generation task which is hard for LLM agents to deal with. In the real analog design process, the analog designers often use the subcircuits instead of the devices. Inspired by this, we propose a novel subcircuit-level SPICE code representation that is built up from subcircuits. For example (as Figure~\ref{subcircuit} shows), devices MM0-2 belong to the same subcircuit XI0 and thus can be simplified to one line. The details of our subcircuit library are shown at the bottom of Figure~\ref{subcircuit}, including one-signal path blocks, two-signal path blocks, capacitors, and resistors. It is important to note that the subcircuit library is summarized from analog design experience and can be extended easily. To summarize a high-quality subcircuit library, we basically refer to subcircuit libraries from related studies~\cite{zhaoDeepReinforcementLearning2022} in electronic design automation (EDA) fields. Moreover, we refer to some analog circuit design books~\cite{analogdesignrazavi} to make minor modifications to the subcircuit library aiming to be more practical.

\subsection{Design Task Decomposition}
\label{prompting}
AnalogXpert mainly leverages CoT and in-context learning. The CoT has two major steps: \textbf{(1) Block Selection.} Based on the previous subcircuit library, the analog topology synthesis task can be performed like the human design process. In this step, LLM agents select the appropriate subcircuits from the library according to the design requirements. \textbf{(2) Block Connection.} With the selected subcircuits, LLM agents then determine the connection relationships and generate the final analog topology. For in-context learning, AnalogXpert selects some design task examples as prompts. Each example contains the design requirements and the corresponding analog topology. The principle of selection is the similarity of design requirements. For a given design requirement, AnalogXpert will give the most similar design task as an example. With the CoT and in-context learning, the basic generation functions can be achieved.         

\subsection{Human experience-based proofreading}
\label{selfrefinement}
\begin{figure*}[htb]
    \centering
    \includegraphics[width=0.9\textwidth]{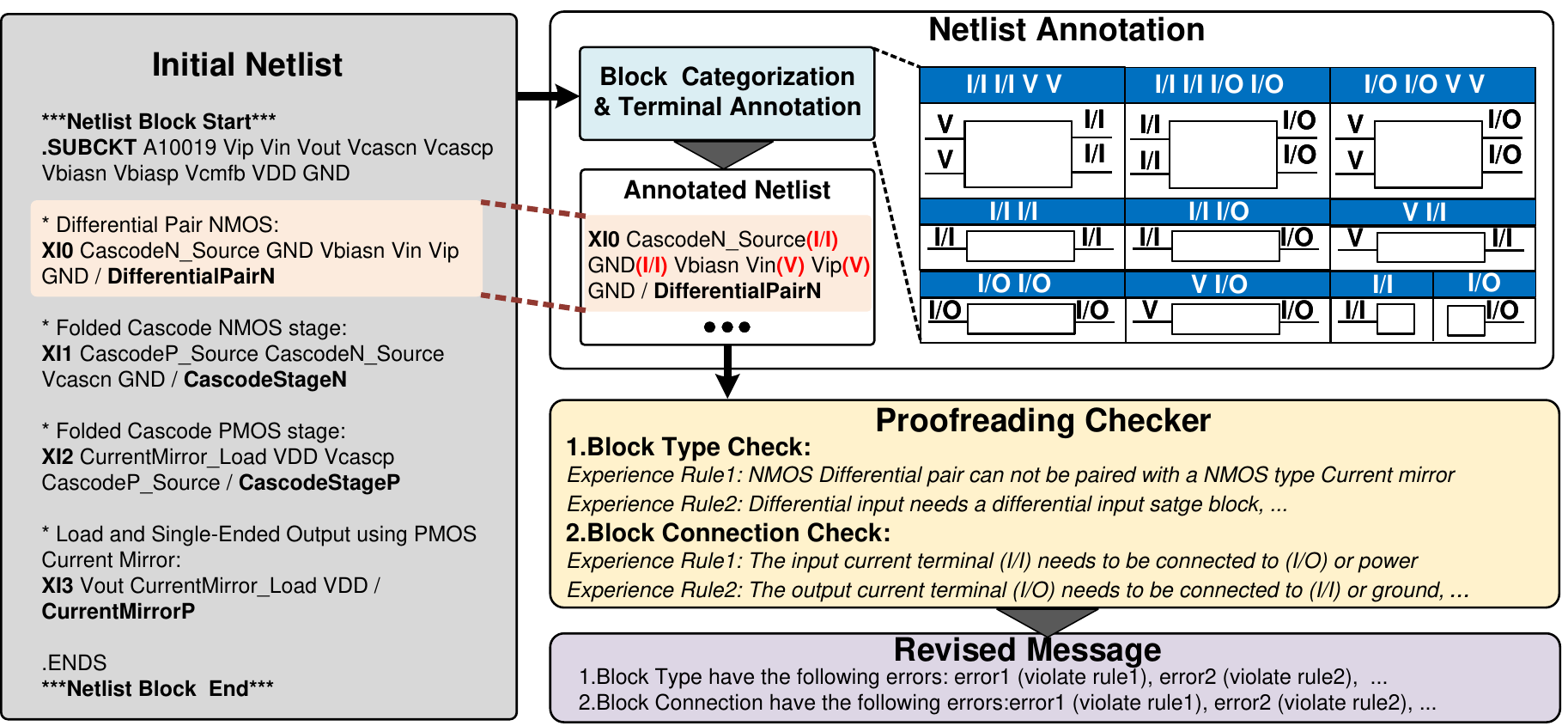}
    \caption{Human experience-based proofreading.}
    \label{self-refine}
    \vspace{-0.2cm}
\end{figure*}

During the generation of analog circuit topology, we expect the LLM agent to follow some design rules. A straightforward idea is that we summarize these rules as text and then use them as prompts. However, when dealing with a series of rules, the prompt gets longer and longer, and it becomes difficult for an LLM agent to fully understand these rules and follow them strictly. Therefore, we propose human experience-based proofreading to get LLM agents out of this dilemma. The basic idea of proofreading is depicted in Figure~\ref{self-refine}. In practice, the initial netlist does not provide enough information for the rule-based checker. These netlists are automatically annotated with the terminal types including current output (I/O), current input (I/I), and voltage (V). In this way, the previous subcircuit library can be summarized into 10 types, making the error-checking process simpler (shown as the right-top part of Figure~\ref{self-refine}). After the annotation, a proofreading checker will detect the corresponding errors, which are mainly categorized into block selection errors and block connection errors. For the accuracy of checking, the proofreading checker is implemented with deterministic programs rather than LLM agents. The detected errors as well as the violated rules make up the final revised message. The LLM agents take the revised message to generate the refined analog topology. AnalogXpert will repeat this process until the design meets the design requirements or reaches the maximum iterations. It is worth noting that not only the current refinement tips are provided, but also the generation and refinement history to avoid making similar mistakes as much as possible.


\section{Experiments}

\begin{table*}[bt]
    \caption{\small{Experimental results including comparisons and ablation study (AnalogXpert is based on GPT-4o).}}
    \label{abliation_study}
    \centering
    \resizebox{\textwidth}{!}{
    \begin{tabular}{c|c|c|c|c|c|c|c|c}
\toprule
TaskId & Task1 & Task2 & Task3 & Task4 (one stage & Task5 (two stage & Task6 (three stage & \multirow{2}{*}{Average} & Real\\
Method & (one stage)  & (two stage) & (three stage) & one component) & one component) & one component) &  & Task\\
\midrule 
\multicolumn{9}{c}{Comparasion} \\
\midrule 

GPT-3.5Turbo & 1/10(10\%) & 0/15(0\%) & 0/20(0\%) & 0/20(0\%) & 0/20(0\%) & 0/15(0\%) & 1/100(1\%) & 1/30(3\%) \\
GPT-4o & 2/10(20\%) & 1/15(6\%) & 0/20(0\%) & 0/20(0\%) & 0/20(0\%) & 0/15(0\%) & 3/100(3\%) & 1/30(3\%) \\
AnalogCoder\cite{laiAnalogCoderAnalogCircuit2024} & 3/10(30\%) & 2/15(13\%) & 0/20(0\%) & 2/20(10\%) & 1/20(5\%) & 0/15(0\%) & 8/100(8\%) & 2/30(6\%) \\
GPT-3.5Turbo+Ours &  5/20(25\%) & 41/125(33\%) & 156/625(25\%) & 60/150(40\%) & 234/750(31\%) & 66/330(20\%) & 562/2000(28\%) & 0/30(0\%) \\
\textbf{AnalogXpert (10R)} & \textbf{16/20(80\%)} & \textbf{53/125(42\%)} & \textbf{221/625(35\%)} & \textbf{86/150(57\%)} & \textbf{327/750(44\%)} & \textbf{98/330(30\%)} & \textbf{801/2000(40\%)} & \textbf{7/30(23\%)} \\
\midrule 
\multicolumn{9}{c}{Ablation Study} \\
\midrule 
WoT CoT\&In context  & 7/20(35\%) & 31/125(25\%) & 162/625(26\%) & 68/150(45\%) & 266/750(35\%) & 95/330(29\%) & 629/2000(31\%) & -- \\
WoT Proofreading & 4/20(20\%) & 6/125(5\%) & 29/625(5\%) & 31/150(21\%) & 64/750(9\%) & 15/330(5\%) & 149/2000(7\%) & -- \\
AnalogXpert (1R) & 7/20(35\%) & 10/125(8\%) & 48/625(8\%) & 32/150(21\%) & 110/750(15\%) & 19/330(6\%) & 226/2000(11\%) & -- \\
AnalogXpert (5R) & 15/20(75\%) & 39/125(31\%) & 122/625(19\%) & 66/150(44\%) & 224/750(30\%) & 54/330(16\%) & 520/2000(26\%) & -- \\
\bottomrule
\end{tabular}
}
\vspace{-0.5cm}
\end{table*}
\textbf{Baseline.} 
We compare AnalogXpert (based on GPT-4o) with pure GPT-3.5, and GPT-4o, which are advanced models with strong code generation ability and rich cross-disciplinary knowledge. The GPT-3.5 and GPT-4o prompts contain only basic task descriptions and do not contain any CoT descriptions or subcircuit library information. We also make some adaptation adjustments to allow Analogcoder~\cite{laiAnalogCoderAnalogCircuit2024} to compare with AnalogXpert.
For other work related to topology synthesis, as they handle different problems with us, the comparison with them is infeasible and thus is excluded.
Specifically, AnalogXpert tackles concrete structure design requirements while most related work explores ambiguous design specifications~\cite{dongCktGNNCircuitGraph2024a,zhaoDeepReinforcementLearning2022,liuLADACLargeLanguage2024,changLaMAGICLanguageModelbasedTopology,chen24artisan}. 
Besides, AnalogXpert directly works on real-world device-level models while the related work focuses on ideal model~\cite{dongCktGNNCircuitGraph2024a,changLaMAGICLanguageModelbasedTopology,chen24artisan}. Therefore, we make the comparison with AnalogCoder~\cite{laiAnalogCoderAnalogCircuit2024}, pure GPT-3.5, and GPT-4o.


\textbf{Benchmark.} We construct two benchmarks for the final evaluation, including a real data benchmark and a synthetic data benchmark. The real data benchmark is collected from a commercial tool named AnalogDesignToolbox~\cite{ADTtool}. We select the most representative 30 analog topologies as the real data benchmark. The synthetic data benchmark is built by a random generation Python code leveraging the subcircuit library. Each synthetic data consists of four parts, the stage number, the input blocks, other given blocks, and the maximum number of blocks. The task of generation on synthetic data is selecting some subcircuit blocks based on the input blocks and other given blocks to form the final circuit design. The total number of used blocks should be less than the maximum number of blocks. The stage is set from one to three, the input blocks and other blocks are randomly selected from the proposed subcircuit library, and the maximum number of blocks is randomly selected from a range related to the stage number(e.g. for one stage the range is 2-5). Finally, we generate 2k synthetic data with totally different structure design requirements. 

\begin{figure*}[tb]
    \centering
    \includegraphics[width=0.88\textwidth]{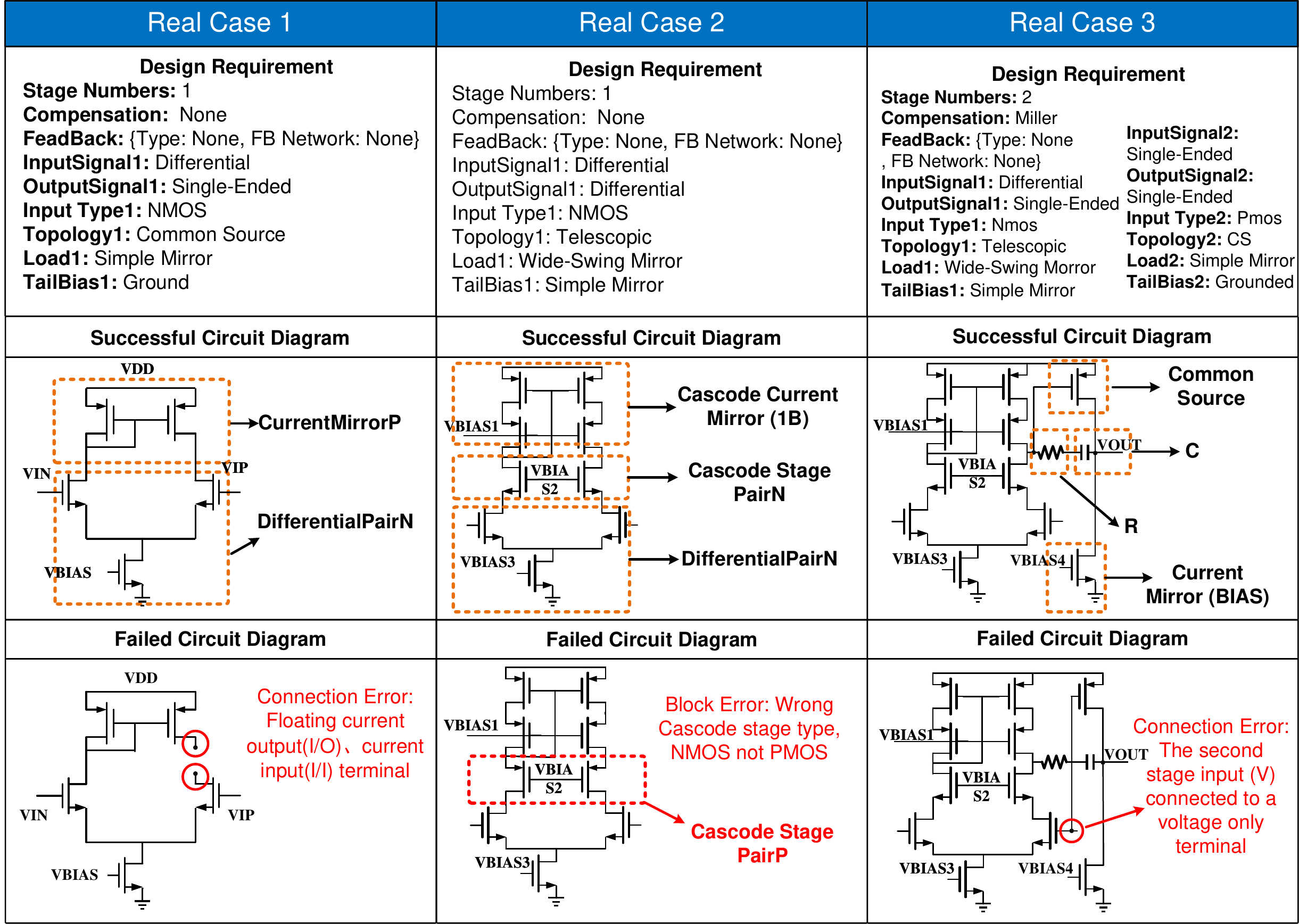}
    \caption{Three real cases with successful circuit diagrams and failed circuit diagrams.}
    \label{result}
   \vspace{-0.5cm}
\end{figure*}

\textbf{Metrics.} The metric of AnalogXpert is whether the LLM agent can generate the correct analog topology for the given design requirements in one trial. For real data, the correctness can only be determined if all blocks and connections exactly match the design requirements. Such strict correctness requires humans to check the analog topology results directly. For synthetic data, the block selection and connection have some basic rules to follow. If a generated circuit topology does not violate these rules, it will be determined as correct. Such correctness can be checked by an automatic program. 
Tested on a certain number of cases, the correct ratio can be obtained. The correct ratio can directly reflect the ability of the LLM agent to generate the required analog topology.

\textbf{Main Results.} In the main experiment (as Table~\ref{abliation_study} shows), each method conducts seven different tasks including six tasks on synthetic data and one task on the real data. The design requirements of Task 1-3 have different input stage numbers (1-3) without any given blocks. The design requirements of Task 4-6 have different input stage numbers (1-3) with an extra given block. The pure GPT-3.5 Turbo, GPT-4o, and AnalogCoder~\cite{laiAnalogCoderAnalogCircuit2024} are only tested on one hundred synthetic data because the automatic check program can not be performed without the subcircuit library-based representation. Thus, the results are checked by humans. On synthetic data Pure GPT-4o and pure GPT-3.5 Turbo can only achieve a correct ratio of 3\% and 1\%, respectively. AnalogCoder~\cite{laiAnalogCoderAnalogCircuit2024} achieves 8\% and 6\% on synthetic data and real data, respectively, which is much lower than AnalogXpert 40\% and 23\%. Such experimental results demonstrate the importance of the proposed subcircuit library-based representation method. We can also observe that AnalogXpert(40\%) outperforms GPT-3.5Turbo+Ours(28\%) on synthetic data. This result indicates that the proposed framework needs models with sufficient comprehension to generate the topology more accurately. On real data, only AnalogXpert achieves a 23\% correct ratio, other methods have near-zero correct ratios. The failure of GPT-3.5Turbo+Ours on real data is due to poor model comprehension and a significant increase in task difficulty. The experimental results on both benchmarks demonstrate the effectiveness of the proposed AnalogXpert in dealing with complex connection relationships and real-world topology synthesis problems.

\textbf{Ablation Study.} Ablation experiments are not performed on the real task because the limited number of real tasks can not reflect statistical trends. We perform the ablation study on each component of AnalogXpert except for the subcircuit library-based representation. Subcircuit library-based representation is the foundation of CoT and proofreading and the framework will become pure GPT-4o without it. The experimental results indicate that CoT \& In-context learning has less impact on performance compared to the proofreading strategy. We also conduct the experiments with different proofreading rounds including one round, five rounds, and ten rounds. We use ten-round proofreading in the proposed AnalogXpert. The experimental results are also consistent with the intuition that the correct ratio increases as the number of proofreading rounds increases, proving the effectiveness of this strategy.

\textbf{Visulization.} Three visualized results of the real data are shown in Figure~\ref{result}. Each case shows the failed circuit design in the generation process and is then corrected by AnalogXpert during the proofreading step. The failure reason in case 1 is the connection error due to the floating current input and output terminal, which is not allowed. In real case 2, the AnalogXpert makes a mistake in the selection of subcircuits. The cascode stage should be N-type, but AnalogXpert first selects the P-type. Real case 3 is a two-stage amplifier, the AnalogXpert connects the input of the second stage to the input of the first stage, which destroys the structure of the two stages.

 












\section{Conclusion}

In this work, we propose AnalogXpert, a powerful analog topology synthesis tool based on training-free LLMs. AnalogXpert leverages the subcircuit-based circuit representation, CoT\&in-context learning, and human experience-based proofreading to imitate the human design process and improve the design accuracy. AnalogXpert achieves 40\% and 23\% success rates on the synthetic and real datasets, respectively, which is better than those of AnalogCoder (8\%,6\%).



{
\small
\bibliographystyle{IEEEtran}
\bibliography{./AnalogXpert.bib}
}

\end{document}